\documentclass[aps,prb,twocolumn, superscriptaddress,showpacs,letterpaper]{revtex4}
\usepackage{graphicx}
\usepackage{psfrag}
\usepackage{subfigure}
\usepackage{verbatim}
\usepackage{color}
\usepackage{upgreek}
\usepackage{amsmath,amssymb}
\usepackage[pdfauthor={Posen Transtrum Catelani Liepe Sethna},
            pdftitle={Shielding superconductors with thin films}]{hyperref}
\hypersetup{pdfstartview={XYZ null null 1.09}}

\newcommand{\bA}{\mathbf{A}}
\newcommand{\bv}{\mathbf{v}}
\newcommand{\I}{{\mathrm{i}}}

\begin{document}


\title{Shielding superconductors with thin films}



\author{Sam Posen}
\affiliation{LEPP, Physics Department, Newman Laboratory, Cornell University, Ithaca, NY 14853-2501}
  \altaffiliation[Now at ]{Fermi National Accelerator Laboratory, Batavia, Illinois, 60510, USA.}
\author{Mark K. Transtrum}
\affiliation{Department of Physics and Astronomy,
Brigham Young University, Provo, UT 84602}
\author{Gianluigi Catelani}
\affiliation{Forschungszentrum J{\"u}lich, Peter Gr{\"u}nberg Institut
(PGI-2), 52425 J{\"u}lich, Germany}
\author{Matthias U. Liepe}
\affiliation{LEPP, Physics Department, Newman Laboratory, Cornell University, Ithaca, NY 14853-2501}
\author{James P.~Sethna}
\affiliation{LASSP, Physics Department, Clark Hall, Cornell University, Ithaca, NY 14853-2501}


\date{\today}

\begin{abstract}
Determining the optimal arrangement of superconducting layers to withstand large amplitude AC magnetic fields is important for certain applications such as superconducting radiofrequency cavities. In this paper, we evaluate the shielding potential of the superconducting film/insulating film/superconductor (SIS') structure, a configuration that could provide benefits in screening large AC magnetic fields. After establishing that for high frequency magnetic fields, flux penetration must be avoided, the superheating field of the structure is calculated in the London limit both numerically and, for thin films, analytically. For intermediate film thicknesses and realistic material parameters  we also solve numerically the Ginzburg-Landau equations. It is shown that a small enhancement of the superheating field is possible, on the order of a few percent, for the SIS' structure relative to a bulk superconductor of the film material, if the materials and thicknesses are chosen appropriately.
\end{abstract}



\maketitle

\section{Introduction}

Can one engineer a better superconducting magnetic shield? How can one optimally arrange materials to maintain complete flux exclusion from a region, and what is the maximum external field that can be screened? It has long been known that superconducting films of width $d$ smaller than the London magnetic penetration depth $\lambda$ can remain superconducting at much higher magnetic fields than bulk samples,\cite{Tinkham} so it has been proposed that films could be used to shield bulk superconductors.\cite{Gurevich2006} In this paper, we investigate the shielding properties of the film/insulator/bulk (SIS') structure and compare to the single superconducting slab. The focus here is on AC rather than DC shielding; the latter has already been studied extensively.\cite{Pavese1998,DenisPhD,claycomb1999,Obi2004,plechacek1996}

Superconducting radio-frequency (SRF) cavities are an example of an application in which shielding of large-amplitude high frequency magnetic fields is required. This technology underlies particle accelerators used in high-energy physics, nuclear physics, neutron sources, and X-ray light sources. The large AC accelerating electric field of these cavities induces a correspondingly large magnetic field.  If the magnetic field exceeds the flux penetration field of the material, it causes a quench in the cavity. If SIS' structures could enhance the flux penetration field relative to that of a bulk superconductor, it could allow these cavities to achieve higher accelerating fields.\footnote{With flux penetration field we mean the maximum value of the magnetic field just outside the film such that no flux enters into the bulk.}
In this paper, we examine the superheating fields $B_{sh}$ of these structures, where flux penetration would occur in defect-free superconductors.
In fact, part of the motivation for this work is that there has been significant confusion in the SRF community regarding the maximum fields that SIS' structures can screen; we hope that this study will clarify the screening mechanism and its limitations. Our calculations show modest shielding gains for SIS' heterolaminates compared to bulk superconductors. The SIS' structure may provide benefits in other ways for realistic materials with surface defects,\cite{Gurevich2015} but considering those benefits is beyond the scope of the present work.

The paper is organized as follows: we start our analysis by arguing in Sec.~\ref{sec:fe} that for an SIS' structure, a significant enhancement of the flux penetration field could be achieved only if a significant gradient in the phase of the order parameter $\nabla\phi$ can be established across the film shielding the bulk. Since this would result in a level of dissipation that is likely unmanageable, we restrict our analysis to fields below $B_{sh}$, where both film and bulk superconductor are in a (meta)stable regime, with no phase gradient across the film. In Sec.~\ref{sec:London}, numerical calculations are performed in the London limit. The thin film regime is examined in Sec.~\ref{sec:thin} with an analytical Ginzburg-Landau approach. In Sec.~\ref{sec:numGL}, the results are extended to films of intermediate thicknesses via a full numerical Ginzburg-Landau analysis. We summarize our work in Sec.~\ref{sec:summary}.

\section{Flux Exclusion}
\label{sec:fe}

The fundamental link between superconducting order and magnetism is the fact that the free energy and properties of the system are governed not by the gradients $\nabla \psi$ of the superconducting order, but by a `covariant' derivative $D\psi = (\nabla - e^* \I \bA / \hbar) \psi$, where $e^*=2e$ is the Cooper pair charge and $\bA$ is the magnetic vector potential. If we write the complex superconducting order parameter in terms of two real fields as $\psi = |\psi| \exp(\I \phi)$, the covariant derivative becomes
\begin{align}
D\psi &= \left[\nabla |\psi|
	+ \I |\psi| (\nabla \phi - e^* \bA / \hbar) \right] \exp(\I \phi) \\
      &= \left[\nabla |\psi|
	+ \I |\psi| (m^* \bv_s /\hbar) \right] \exp(\I \phi),
\end{align}
where the gauge-invariant combination
\begin{equation}
(\hbar/m^*)(\nabla \phi - e^* \bA/\hbar) = \bv_s
\label{eq:vs}
\end{equation}
is called the {\em supercurrent velocity}. Magnetic fields cause ``stress'' in superconductors indirectly through $\bA$, which induces screening supercurrents. Due to these supercurrents, a weak magnetic field exponentially decays inside a superconductor over the penetration depth $\lambda$. As a crude approximation, the superconductor can support a certain maximum stress, characterized by a maximum superfluid velocity $\bv_s^{max}$. The superconductor can only screen $A$ values larger than $v_s^{max}m^*/e^*$ if it passes vortex lines through its boundary. For example, if a vortex line is passed through a hollow superconducting cylinder in a parallel external field, this will bring flux inside the cylinder and $\phi$ will wind by a factor of $2\pi$, lowering the stress in the superconductor.

\begin{figure}[htbp]
\begin{center}
\includegraphics[width=0.48\textwidth,angle=0]{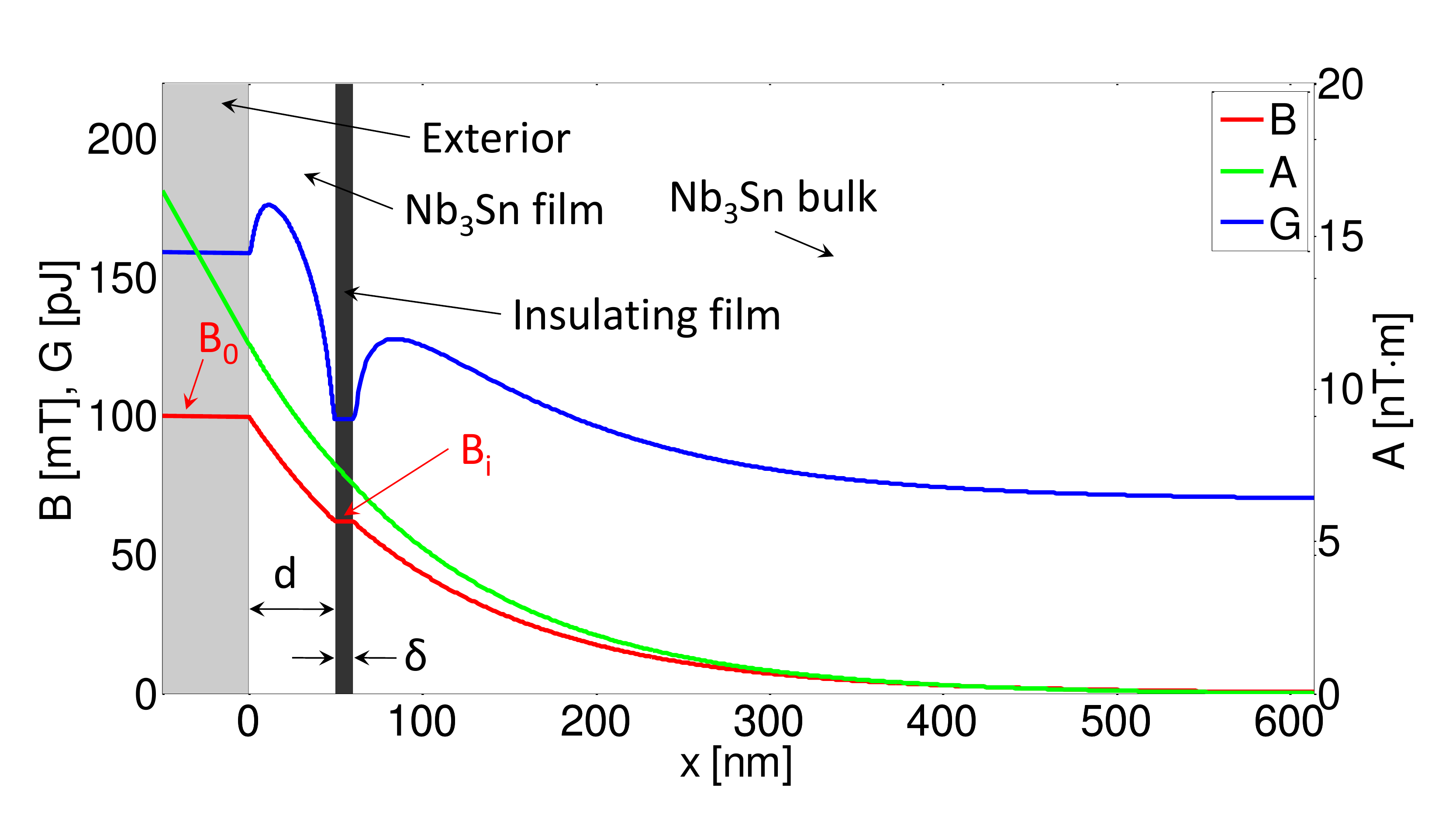}
\end{center}
\caption{Example of a SIS' structure. The amplitudes of the magnetic field, the vector potential, and the Gibbs free energy are plotted as a function of distance into the structure.}
\label{fig:geometry}
\end{figure}

Now let us consider a single thin superconducting film separated from a bulk superconductor by a thin insulator, shown in Figure \ref{fig:geometry}. In a `thin' superconductor of thickness $d\ll\lambda$ the critical fields are enhanced; for example, for the parallel thermodynamic critical field we have $H_{c\parallel}=2\sqrt{6}\frac{H_c \lambda}{d}$.\cite{Tinkham} The Meissner state requires $\bA\to 0$ deep in the bulk, and $\bA$ is continuous across the insulating gap. Therefore, the vector potential at the film surface is tied to that of the bulk superconductor surface; however, the insulating gap offers the opportunity to decouple the phase gradient across the film from that in the bulk. If many vortex lines pass through the film, the superconducting film could be relatively unstressed, supplementing the native superheating field of the film material.

In DC, it should be possible to screen a bulk from very large fields using a compound film with many layers of alternating thin superconducting and insulating films with magnetic flux trapped between each of them. However, in AC applications, filling the insulators with magnetic flux demands the transfer of $\nabla\phi/\pi$ fluxoids per unit length across the screening film in each cycle. As they pass through the film, the vortices would experience strongly dissipative drag,\cite{Gurevich2006} generating levels of heating that are likely unmanageable for most applications.\cite{SRF13}

As we are focusing on high frequency applications, we impose the restriction that flux must never pass into the superconducting regions. With this restriction, the SIS' structure would offer an advantage over a single thick superconducting slab if it could withstand higher magnetic fields without flux penetration. The flux-free state is only intrinsically stable below $B_{c1}$, the lower critical field. However, there is good evidence that real materials can withstand high frequency fields well above $B_{c1}$. \cite{liepe2013,Valles2014} In this metastable regime, an energy barrier prevents flux from penetrating, a barrier that is reduced to zero at $B_{sh}$ for a defect-free material (thermal fluctuations at cryogenic temperatures are much smaller than the condensation energy, so they cannot create excitations above the barrier). $B_{sh}$ is the ultimate AC magnetic limit; this is especially important for SIS' films, as they are always in the metastable state.\cite{SRF13} We will use $B_{sh}^{SIS'}$ to denote the maximum metastable field of a SIS' structure to distinguish it from the superheating field of the bulk material, $B_{sh,b}$, and the bulk superheating field of the film material, $B_{sh,f}$ (i.e. the value it would have if it were not a thin film). In the next three sections we present and compare three approaches to evaluate $B_{sh}^{SIS'}$.

\section{Superheating field in the London limit}
\label{sec:London}

To make a rough estimate of the superheating field of the SIS' structure, we consider the Gibbs free energy $\mathcal{G}$ in the London limit; that is, we assume that both film and bulk superconductors are strongly type II materials, with penetration depths much longer than coherence lengths. We denote by $\lambda_f$ the film's material penetration depth and by $\xi_f$ its coherence length. The thickness $d$ of the film is assumed to be much larger than $\xi_f$; in particular, for the vortex core to be accommodated in the film one needs $d\gtrsim 1.8 \xi_f$\cite{SaintJames1965,Fink1969}. The film is separated from a bulk superconductor with with penetration depth $\lambda_b$ by an insulating film of thickness $\delta$. The superconducting film is screening the bulk from a parallel magnetic field with amplitude $B_0$. The screened field between the film and the bulk has amplitude $B_i$. In our geometry, the $x$-axis is perpendicular to the film, pointing into it, with origin at the interface with the exterior. The $z$-axis is aligned with the magnetic field.

The Gibbs free energy of a vortex in a superconductor can be determined from the value of two magnetic fields evaluated at the vortex location $r_0$: the Meissner-screened external field $B_M$ and the field generated by the vortex in the film $B_V$:\cite{Stejic1994}
\begin{equation}
\mathcal{G}=\frac{\phi_0}{\mu_0}\left(B_V(r_0)/2+B_M(r_0)\right) \, ,
\label{eq:gibbs}
\end{equation}
where $\phi_0$ is the flux quantum and $\mu_0$ the magnetic constant. The field $B_M$ can be found by minimizing the free energy in the structure when no vortex is present; we remind that in the London limit the Meissner field in the bulk superconductor decays exponentially and hence it equals $B_i e^{-(x-(d+\delta))/\lambda_b}$. This procedure gives:
\begin{equation}
B_M=\frac{B_0+B_i}{2}\frac{\cosh{\frac{x-d/2}{\lambda_f}}}{\cosh{\frac{d}{2\lambda_f}}}-\frac{B_0-B_i}{2}\frac{\sinh{\frac{x-d/2}{\lambda_f}}}{\sinh{\frac{d}{2\lambda_f}}}\, ,
\label{eq:Bm}
\end{equation}
where $B_i$ is given by
\begin{equation}
B_i=B_0\left[\frac{\delta+\lambda_b}{\lambda_f}\sinh{\frac{d}{\lambda_f}}+\cosh{\frac{d}{\lambda_f}}\right]^{ -1} \, .
\label{eq:Bi}
\end{equation}

\begin{table}[!tb]
\begin{center}
\begin{tabular}{| c | r | r | r | r | r |}
\hline Material & $\lambda$ [nm] & $\xi$ [nm] & $B_{c1}$ [T] & $B_{c}$ [T] & $B_{sh}$ [T] \\ \hline\hline
    Nb & 40 & 27 & 0.13 & 0.21 & 0.25 \\ \hline
    Nb$_3$Sn & 111 & 4.2 & 0.042 & 0.50 & 0.42 \\ \hline
    NbN & 375 & 2.9 & 0.006 & 0.21 & 0.17\\ \hline
    MgB$_2$ & 185 & 4.9 & 0.017 & 0.26 & 0.21 \\ \hline
\end{tabular}
\caption{{\bf Materials parameters} of niobium and three promising alternative SRF materials. The penetration depth $\lambda$ is calculated using Eq.~3.131 in Ref.~\onlinecite{Tinkham2004}. The correlation length $\xi$ is calculated using the equations in Ref.~\onlinecite{Orlando1979}. For Nb a RRR of 100 was assumed. For MgB$_2$, $\lambda$ and $\xi$ are not calculated, as the experimental values are given in the reference. For calculations, $B_c=\phi_0/(2\sqrt{2}\pi\xi\lambda)$ is used.\cite{Tinkham2004} $B_{c1}$ for Nb found from power law fit to numerically computed data from Ref.~\onlinecite{Hein1999} and \onlinecite{Harden1963} and for strongly type II materials is found from Eq.~5.18 in Ref.~\onlinecite{Tinkham2004}. $B_{sh}$ is calculated using $B_{sh}\simeq B_c\left(\frac{\sqrt{20}}{6}+\frac{0.5448}{\sqrt{\kappa}}\right)$ from Ref.~\onlinecite{Transtrum2011}. Nb data from Ref.~\onlinecite{Maxfield1965}, Nb$_3$Sn data from Ref.~\onlinecite{Hein2001}, NbN data from Ref.~\onlinecite{Oates1991}, and MgB$_2$ data from Ref.~\onlinecite{Wang2001}. Note that the two gap nature of MgB$_2$ may require more careful analysis than is performed here.}
\label{tab:MaterialsParameters}
\end{center}
\end{table}

\begin{figure}[htbp]
\begin{center}
\includegraphics[width=0.48\textwidth,angle=0]{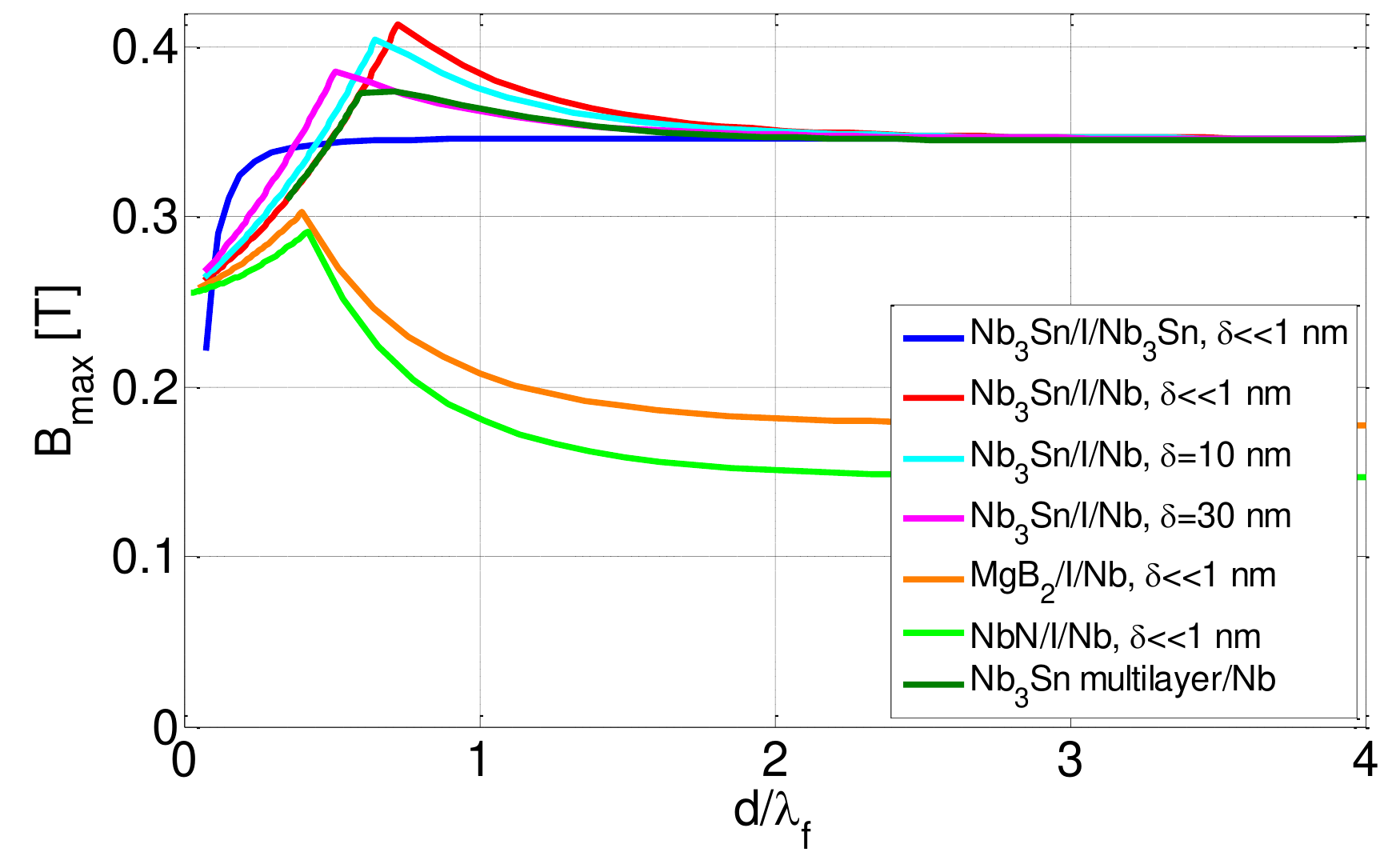}
\end{center}
\caption{Maximum field below $B_{sh}$ of both the film and the bulk as a function of film thickness for various film materials in a SIS structure with Nb. The effect of varying the insulator thickness $\delta$ is shown for the Nb$_3$Sn film, as is the effect of splitting the film thickness $d$ over 5 equally thick multilayers with thin separating insulators. All calculations done in the London limit.}
\label{fig:BshvsdMulti1}
\end{figure}

Explicit formulas for $B_V$ are available for thin ($d\ll\lambda_f$) and thick ($d\gg\lambda_f$) films.\cite{Stejic1994} To study the full range of thicknesses, we use the more general expression of Ref.~\onlinecite{Shmidt1972} (this expression assumes $r_0=(x_0,0)$):
\begin{equation}
B_V=\frac{2\phi_0}{\lambda^2d}\sum\limits_{n=1}^\infty\int\limits_{-\infty}^\infty \frac{dk}{2\pi}e^{iky}\frac{\sin(\pi n x/d)\sin(\pi n x_0/d)}{k^2+(\pi n/d)^2+1/\lambda^2}
\label{eq:BV}
\end{equation}
Equations~(\ref{eq:Bm})-(\ref{eq:BV}) give the fields in the structure, and Eq.~(\ref{eq:gibbs}) gives the Gibbs free energy as shown in Fig.~\ref{fig:geometry}. The barrier to flux penetration is due to the positive slope of $\mathcal{G}$ inside the superconducting regions near the interfaces. We can find $B_{sh}^{SIS'}$ by finding the field at which the barrier is reduced to zero in any of the superconductors\footnote{Niobium has fairly low kappa, so the London limit does not apply. In these calculations, the limit used for bulk niobium is when the surface reaches $B_{sh}$ from Table \ref{tab:MaterialsParameters}. For all other materials, the slope of the Gibbs free energy is used as the criterion for flux penetration as described.}. In Fig.~\ref{fig:BshvsdMulti1}, $B_{sh}^{SIS'}$ is plotted as a function of superconducting film thickness for various SIS' structures. Various insulator thicknesses are considered, including the thin layer limit, for illustrative purposes as it gives the highest fields. The materials analyzed are those that are promising for SRF cavities, with properties given in Table~\ref{tab:MaterialsParameters}.

\begin{figure}[htbp]
\begin{center}
\includegraphics[width=0.48\textwidth,angle=0]{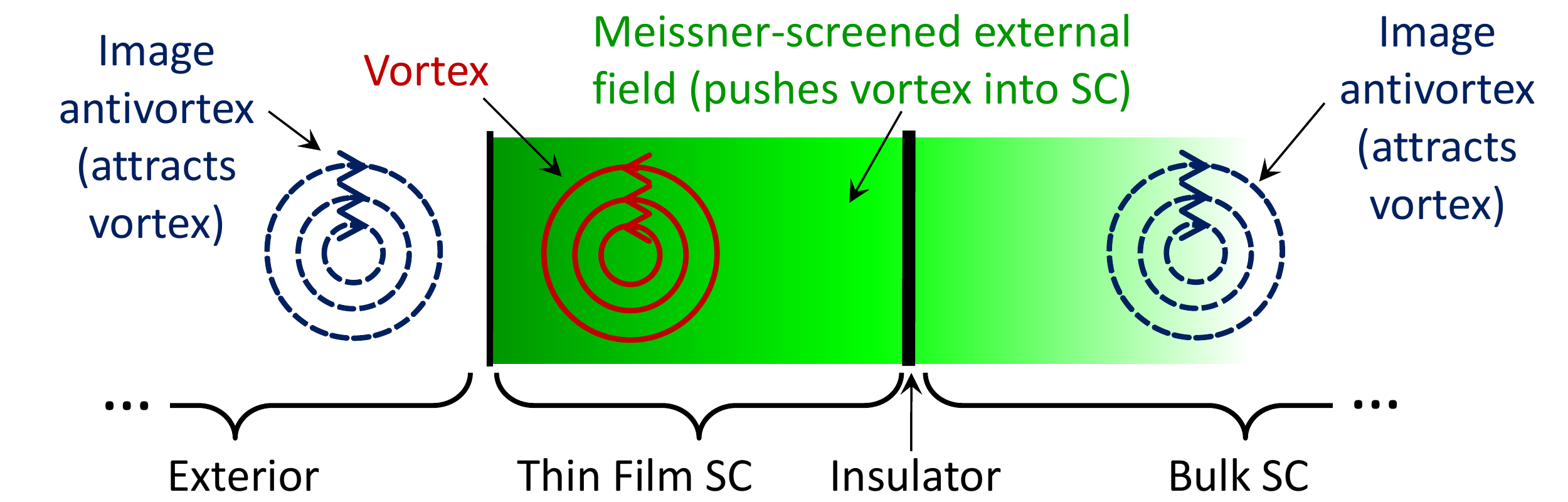}
\end{center}
\caption{Forces on a vortex in a homolaminate. As the film is made thinner, the image antivortex to the right of the film has a stronger pull on the vortex, lowering the barrier to vortex penetration.}
\label{fig:vortex}
\end{figure}

The structures plotted in Figure \ref{fig:BshvsdMulti1} can be divided into two types: homolaminates, in which the film is the same material as the bulk, and heterolaminates, in which they are different. Calculations show that for a homolaminate like Nb$_3$Sn/insulator/Nb$_3$Sn, the film is the weak point: it always reaches its $B_{sh}$ before the bulk, and the thinner the film, the lower its $B_{sh}$. Homolaminates with films that are so thick that they behave like a bulk superconductor have the highest $B_{sh}^{SIS'}$. To better understand this, consider the magnetic forces on a vortex [which can be derived from Eq.~(\ref{eq:gibbs})]. The boundary condition imposed by $B_V$ can be satisfied by an image antivortex outside of the boundary, which creates a force that pulls the vortex out of the film.\cite{Bean1964} As the film thickness is reduced, $B_M$ remains approximately unchanged, but the image antivortex on the insulator side of the film used to satisfy $B_V$ has a stronger effect, as shown in Figure~\ref{fig:vortex}. This lowers the barrier to penetration.

\begin{figure}[htbp]
\begin{center}
\includegraphics[width=0.44\textwidth,angle=0]{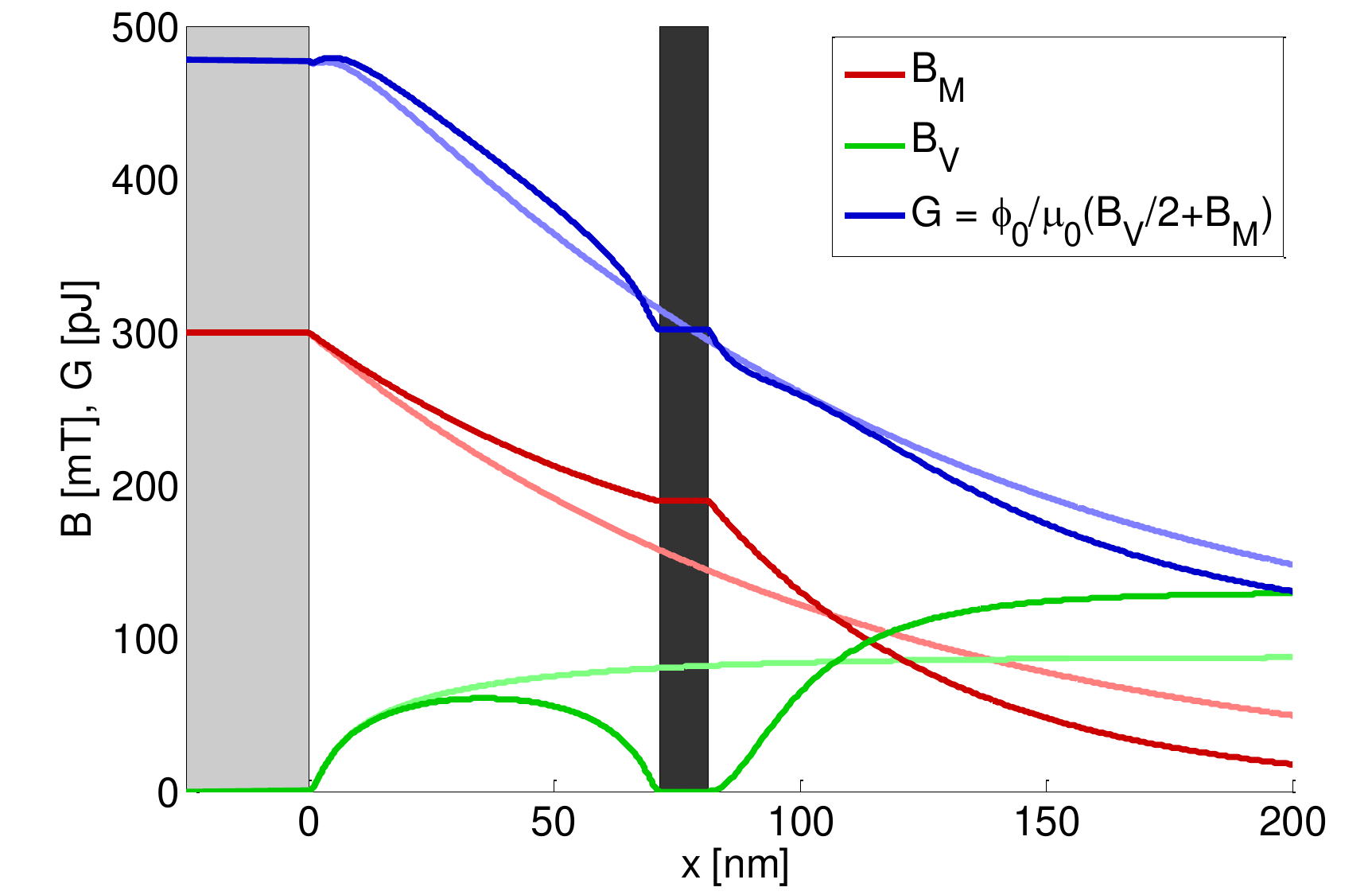}
\end{center}
\caption[Comparison of a SIS' to a bulk film]{Comparison of a SIS' film with near-optimal parameters (dark curves) to a bulk film (light curves). The slower decay of $B_M$ in the large-$\lambda$ thin film influences $\mathcal{G}$, bolstering the barrier to flux penetration. Note that the dark shaded region representing the insulating region of the SIS' can be ignored for the bulk film.}
\label{fig:compareBM}
\end{figure}

The differing penetration depths in the layers of a heterolaminate cause it to behave differently than a homolaminate. Here we consider structures in which the bulk has a smaller penetration depth than the film. For such structures, if the film is very thin, it does not provide much screening for the bulk, and $B_i$ reaches the bulk's $B_{sh}$ before the thin film barrier disappears. As with a homolaminate, a very thick film behaves like a bulk, and reaches that material's bulk $B_{sh}$ while $B_i$ is still relatively small. However, between these two extremes, there is a situation in which the film provides some screening, so that $B_i$ is large but still smaller than $B_0$. In this case, a benefit can be realized -- the small penetration depth of the material in the bulk causes $B_i$ to be larger than with the exponential decay expected for a thick film [Eq.~(\ref{eq:Bi})]. This in turn reduces the magnitude of the negative gradient in $B_M$, bolstering the barrier to flux penetration (Eqn.~\ref{eq:gibbs}). This increase in the barrier is depicted in Fig.~\ref{fig:compareBM}. The dark curves show $B_M$, $B_V$, and $\mathcal{G}$ for a Nb$_3$Sn thin film/insulator/Nb bulk SIS structure with 10 nm thick insulator and $d/\lambda_f=0.64$ (the peak of the cyan curve in Fig.~\ref{fig:BshvsdMulti1}). The light curves show calculations for a bulk Nb$_3$Sn film (for this case, the dark shaded region representing the insulator does not apply). In this example, $B_0=300$ mT. The Gibbs free energy of the SIS' structure is still sharply peaked, showing a relatively robust energy barrier, but that of the bulk film is almost flat, showing that flux penetration is likely to occur at slightly higher fields.

The impact of this is a modest increase in $B_{sh}^{SIS'}$ for these structures compared to the bulk value of the film material. However, the range of film thicknesses over which the increase is appreciable ($\gtrsim$ few $\%$) is relatively small, and the gain decreases as the thickness of the insulating layer increases.


\section{Thin films in the Ginzburg-Landau approach}
\label{sec:thin}

Calculating $B_{sh}$ using London theory, as done in the previous section, fails to take into account 2D instabilities in the order parameter, therefore overestimating $B_{sh}$ in many circumstances. The problem of calculating $B_{sh}$ for bulk samples while taking into account 2D instabilities has a long history (see, \textit{e.g.}, Ref.~\onlinecite{Transtrum2011}) and has mostly been tackled in the Ginzburg-Landau (GL) framework. Only recently calculations beyond GL theory were performed;\cite{Catelani2008,Lin2012} they showed that while the GL results cannot be trusted quantitatively at low temperatures, they give a qualitatively correct estimate. Therefore, for simplicity we restrict ourselves to GL theory even in the low-temperature regime where its quantitative predictions are not exact.

The approach we use to find $B_{sh}$ is described in detail in Ref.~\onlinecite{Transtrum2011}: we first extremize the GL free energy, a functional of the spatially dependent order parameter $\psi$ and superfluid velocity $\bv_s$, and then study its stability against small perturbation of these functions. In the present case, the GL free energy is the sum of the contributions for the bulk and the film. The boundary conditions are the usual ones for the order parameter (vanishing of its derivatives at all surfaces); the superfluid velocity vanishes deep into the bulk, and its derivative at the external film surface is proportional to the applied magnetic field. Similarly, the field between the film and the bulk is proportional to the derivatives of the superfluid velocities at the two surfaces. However, this internal field is not externally imposed, but must be calculated consistently with Maxwell equations; this gives the final condition of continuity of the vector potential across all surfaces.
Hence for a very thin insulating barrier the superfluid velocities at the bulk surface and the internal film surface coincide, while the film superfluid velocity would be higher for a thicker insulator.

For an analytical estimate of the SIS' superheating field $B_{sh}^{SIS'}$, we consider the simplest possible case of a strongly type II bulk material (GL parameter $\kappa_{GL} \gg 1$) shielded by an insulator of negligible thickness and a strongly type II thin film with $\xi_f \ll d \ll \lambda_f$.
With a thin film, the difference between the internal field (at the bulk surface) and the applied field (outside the film) is small, and the maximum possible field is reached when the internal field coincides with the bulk superheating field $B_{sh,b}$. Indeed,
within GL theory and at linear order in $d/\lambda_f$, using the boundary conditions discussed above we find [see Appendix~\ref{appendix}]:
\begin{equation}\label{Bsh}\begin{split}
B_{sh}^{SIS'} = B_{sh,b} \Bigg[1+ \sqrt{\frac65}  \frac{\lambda_b}{\lambda_f} \left(1-\frac{v_{s,r}^2}{3}\right)\frac{d}{\lambda_f} \\
+ \frac12\left(1-v_{s,r}^2\right) \left(\frac{d}{\lambda_f}\right)^2 \Bigg]\, ,
\end{split}\end{equation}
where $v_{s,r} = v_{s,b}^{max}/v_{s,f}^{max}$ is the ratio of the maximum superfluid velocities for bulk and film material, respectively. This ratio can be written in terms of critical fields and penetration depths as $v_{s,r} = B_{c,b}\lambda_b/B_{c,f}\lambda_f$, and as a necessary condition for metastability it must satisfy $v_{s,r} <1$: since in the bulk material the superfluid velocity has already reached its maximum possible value at the surface, the film material must be able to support a higher superfluid velocity.  We stress again that for sufficiently thin films (below the critical thickness discussed in the next paragraph), as the applied field becomes larger than $B_{sh}^{SIS'}$, the bulk becomes unstable, while the film is still (meta)stable.
As qualitatively expected, Eq.~(\ref{Bsh}) shows that for better screening a thicker film should be used, and that as the film material penetration depth increases, its screening power decreases. Also, the need of small $v_s^r$ implies that the film material critical field should be sufficiently large, $B_{c,f}>B_{c,b} \lambda_b/\lambda_f$. Interestingly, based on the values reported in Table~\ref{tab:MaterialsParameters}, this condition can be met if using Nb$_3$Sn or MgB$_2$ to shield Nb.

We note, however, that there is in principle a limit on how thick the film can be made: since the supercurrent velocity at the film external surface increases with thickness, if the film is too thick it will become unstable at a field below that predicted by Eq.~(\ref{Bsh}). Within our approximations, we find that the critical thickness for the film to also become unstable at $B_{sh}^{SIS'}$ is $d_c=\lambda_f\sqrt{6/5}(1-v_{s,r})B_{c,f}/B_{c,b}$. We see that the condition $d<d_c$ can severely restrict the maximum film thickness only in the regime $B_{c,f}\ll B_{c,b}$, $\lambda_f \gg \lambda_b$. For the material parameters in Table~\ref{tab:MaterialsParameters} our formula gives $d_c \sim \lambda_f$, but films of this thickness are beyond the approximate analytical treatment that leads to Eq.~(\ref{Bsh}). Therefore, to study the screening properties of films of intermediated thickness, $d\sim \lambda_f$, in the next section we resort to numerical calculations that also account for the finite value of $\kappa_{GL}$.

\section{Films of intermediate thickness}
\label{sec:numGL}

For films of intermediate thickness, numerical techniques are needed to accurately estimate the effective superheating field of SIS structure in Ginzburg-Landau theory.  Here, we follow closely the methods described in Ref.~\onlinecite{Transtrum2011}.

The Ginzburg-Landau equations are solved in each domain separately and then boundary conditions are matched.  In order to improve numerical stability, we implement the boundary conditions as follows: at the film surface the gradient of the order parameter is fixed to zero while the magnitude is allowed to vary, effectively defining the applied magnetic field implicitely in terms of the order parameter.  We also allow the value of the order parameter and the vector potential on the film side of the interface to vary.  On the bulk side of the interface, the gradient of the magnetic field is fixed to zero while its magnitude is allowed to vary.  Infinitely deep in the bulk the order parameter is fixed to one and the vector potential vanishes.  This configuration introduces three parameters for the boundary conditions: the magnitude of the order parameter on either side of the interface and the magnitude of the vector potential on the film side of the interface.  These three parameters are varied until the gradient of the order parameter vanishes on the film side of the interface and both the magnetic vector potential and the magnetic field are continuous at the interface.

\begin{figure}[htbp]
\begin{center}
\includegraphics[width=0.48\textwidth,angle=0]{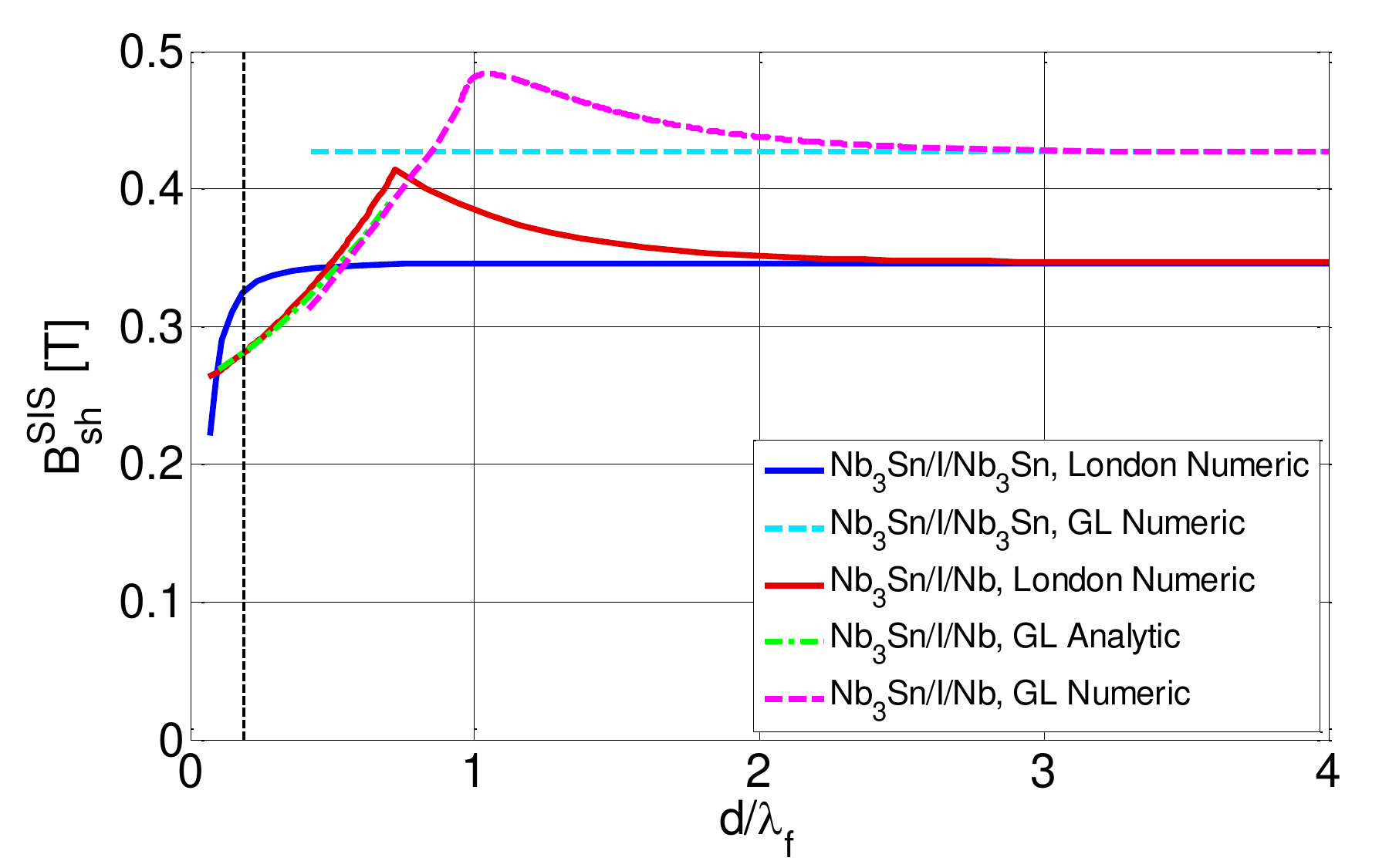}
\end{center}
\caption{$B_{sh}$ in a SIS' structure as a function of film thickness. The insulating layer is assumed to be very thin. London limit calculations are compared to Ginzburg-Landau analytical and numerical calculations. For reference, the dashed vertical line is at position $d \sim 5\xi_f \sim \sqrt{\lambda_f \xi_f}$. We remind that the results of Secs.~\ref{sec:London} (solid lines) and \ref{sec:thin} (dot-dashed line) are valid for films thick compared to $\xi_f$}
\label{fig:BshvsdMulti2}
\end{figure}

Having found a solution, we next solve the eigenvalue problem associated with the stability of the solution to infinitesimal fluctuations of wavenumber $k$ as in Ref.~\onlinecite{Transtrum2011}.  These solutions are also found numerically using boundary conditions similar to those just described.  The magnitude of the applied magnetic field and the wavenumber are then varied simultaneously to identify the least stable fluctuation and the applied magnetic field at which it just becomes unstable (i.e., at which the eigenvalue becomes zero).  In this way we identify the superheating field and the critical wavenumber that characterizes the unstable fluctuations.  These calculations are summarized in Fig.~\ref{fig:BshvsdMulti2} in which we plot $B_{sh}^{SIS'}$ as a function of film thickness for various materials. Note that the dashed lines start from thicknesses of about 50 nm.  For films with thickness less than this, numerical results become increasingly difficult, presumably due to the extremely separated length scales involved.  Interestingly this thickness coincides with approximately twice the depth $\sqrt{\lambda_f \xi_f}$ of the fluctuations\cite{Transtrum2011}, suggesting that interactions between the fluctuations of both films surfaces may become relevant.  Moreover, numerical solutions indicate that at finite $\kappa$, the nature of the of the instability itself may change from 2D to 1D as the thickness decreases.  Although beyond the scope of the present work, these indications deserve further investigation.

The Ginzburg-Landau calculations show good qualitative agreement with the London calculations from Fig.~\ref{fig:BshvsdMulti1}, also shown in this figure. There are some quantitative differences, likely due to the approximations used in the London limit. For instance, the difference in the calculated bulk $B_{sh}$ of the film material, which is approached as the film becomes a few $\lambda_f$ thick, is due to finiteness of $\kappa$. 
For the heterolaminate, in both cases
as the film thickness increases, $B_{sh}^{SIS'}$ shows a peak near $d \sim \lambda_f$, then decreases to the superheating field of the film material as the film becomes very thick. The thickness at which the peak occurs is somewhat smaller for the London limit, but the two plots are otherwise very similar in shape.


\section{Conclusions}
\label{sec:summary}

In this study, we analyzed the magnetic shielding properties of superconductors at high fields and high frequencies. 
To prevent strong vortex dissipation due to drag, the analysis was restricted to a regime where flux penetration is not allowed. The London-limit numerical results were verified against analytical and numerical Ginzburg-Landau calculations. We showed that the SIS' structure can produce a modest enhancement of the maximum screening field compared to a single superconducting slab for certain materials and film thicknesses, see the maxima in Fig.~\ref{fig:BshvsdMulti2}.

\begin{acknowledgments}
Work supported by DOE award number DE-SC0002329, NSF DMR 1312160, and in part by the EU under REA grant agreement CIG-618258.
\end{acknowledgments}

\appendix

\section{Derivation of Eq.~\ref{Bsh}}
\label{appendix}

In the case of large GL parameter $\kappa_{GL} \gg 1$, the calculation of the metastability field, Eq.~\ref{Bsh}, is greatly simplified: for $\kappa_{GL}\to \infty$, the spatial profile of the order parameter is fully determined by that of the superfluid velocity, and the differential equation for the latter is local, albeit non-linear.\cite{Transtrum2011} Indicating with $q_0$ the dimensionless superfluid velocity, for the geometry we are considering it obeys the equation:
\begin{equation}\label{q0eq}
q_0'' = q_0 - q_0^3
\end{equation}
and the metastability condition takes the simple form $q_0^2 < 1/3$.\cite{Transtrum2011} The dimensionful velocity is proportional to $q_0$ multiplied by the critical field and the penetration depth, $v_s \propto B_c \lambda$, and we will not need the proportionality constant in what follows.

For simplicity, in this Appendix we use a coordinate system in which $x$-axis perpendicular to the film has its origin in the middle of the film and measure lengths in units of the film material penetration depth $\lambda_f$. We also take the insulator thickness to be negligible, $\delta =0$, as this gives the highest possible metastable field. As discussed in Secs.~\ref{sec:London} and \ref{sec:thin}, the instability happens at the bulk surface; this fixes the values of the superfluid velocity at the interior surface of the film to be:
\begin{equation}\label{q0bc}
q_0 \left(\frac{d}{2\lambda_f}\right) = -\sqrt{\frac13} v_{s,r}\, , \quad v_{s,r} = \frac{B_{c,b} \lambda_b}{B_{c,f} \lambda_f} \, .
\end{equation}
Clearly, a necessary condition for the metastability of the film is $v_{s,r} < 1$. In fact, since the superfluid velocity at the outer surface is larger, we will further need to check that $q_0^2(-d/2\lambda_f) < 1/3$. In addition to the above boundary condition, we also need the field between film and bulk to coincide with the bulk superheating field:
\begin{equation}\label{bibc}
q_0'\left(\frac{d}{2\lambda_f}\right) = \frac{B_{sh,b}}{\sqrt{2} B_{c,f}} \, .
\end{equation}
The task is now to find the external field at which these two boundary conditions are satisfied,
\begin{equation}\label{bsis}
B_{sh}^{SIS'} = \sqrt{2} B_{c,f} \, q_0'\left(-\frac{d}{2\lambda_f}\right) \, .
\end{equation}

To solve Eq.~\ref{q0eq}, thanks to the assumption $d \ll \lambda_f$, we can proceed by a Taylor expansion of the function $q_0(x)$ near $x=0$:
\begin{equation}
q_0(x) = q_c + b_0 x + b_1 \frac{x^2}{2} + b_2 \frac{x^3}{3} + \ldots
\end{equation}
Substituting the expansion into Eq.~\ref{q0eq} and matching the terms on the two side of the equality we find
\begin{equation}
b_1 = q_c \left(1-q_c^2\right)\, , \quad b_2 = \frac12 b_0 \left(1-3q_c^2\right)\, ,
\end{equation}
showing that only two parameters of the expansion, $q_c$ and $b_0$, are left undetermined and thus can be fixed by the boundary conditions. Moreover, Eq.~\ref{bsis} can be written in the form
\begin{equation}\label{bsis2}
B_{sh}^{SIS'} = B_{sh,b} - \sqrt{2} B_{c,f} q_c (1-q_c^2) \frac{d}{\lambda_f} + {\cal O}\left(\frac{d}{\lambda_f}\right)^3 \, ,
\end{equation}
and hence to calculate $B_{sh}^{SIS'}$ to second order in $d/\lambda_f$ we only need to know $q_c$ to first order. We can therefore use the boundary condition (\ref{bibc}) at lowest order to obtain $b_0 = B_{sh,b}/\sqrt{2} B_{c,f}$ and the boundary condition (\ref{q0bc}) at first order to find
\begin{equation}
q_c =  -\sqrt{\frac13} v_{s,r} - \frac{B_{sh,b}}{\sqrt{2} B_{c,f}} \frac{d}{2\lambda_f} \, .
\end{equation}
Substituting this expression into Eq.~\ref{bsis2} and keeping only terms up to second order we find
\begin{equation}\begin{split}
B_{sh}^{SIS'} = B_{sh,b}\left[1+\frac12 \left(1-v_{s,r}^2\right) \left(\frac{d}{\lambda_f}\right)^2 \right] \\
+ \sqrt{2} B_{c,f}\sqrt{\frac13} v_{s,r} \left(1-\frac{v_{s,r}^2}{3}\right) \frac{d}{\lambda_f} \, .
\end{split}\end{equation}
To put this equation in the form given in Eq.~\ref{Bsh} we use the relationship\cite{Transtrum2011} $B_{sh,b} = \sqrt{5}B_{c,b}/3$ between superheating and critical fields. Finally, by considering at linear order in $d/\lambda_f$ the metastability requirement $q_0(-d/2\lambda_f) > -1/\sqrt{3}$, we obtain the critical thickness $d_c$ reported at the end of Sec.~\ref{sec:thin}.


\bibliography{ThinFilm}

\begin{thebibliography}{27}%
\makeatletter
\providecommand \@ifxundefined [1]{%
 \@ifx{#1\undefined}
}%
\providecommand \@ifnum [1]{%
 \ifnum #1\expandafter \@firstoftwo
 \else \expandafter \@secondoftwo
 \fi
}%
\providecommand \@ifx [1]{%
 \ifx #1\expandafter \@firstoftwo
 \else \expandafter \@secondoftwo
 \fi
}%
\providecommand \natexlab [1]{#1}%
\providecommand \enquote  [1]{``#1''}%
\providecommand \bibnamefont  [1]{#1}%
\providecommand \bibfnamefont [1]{#1}%
\providecommand \citenamefont [1]{#1}%
\providecommand \href@noop [0]{\@secondoftwo}%
\providecommand \href [0]{\begingroup \@sanitize@url \@href}%
\providecommand \@href[1]{\@@startlink{#1}\@@href}%
\providecommand \@@href[1]{\endgroup#1\@@endlink}%
\providecommand \@sanitize@url [0]{\catcode `\\12\catcode `\$12\catcode
  `\&12\catcode `\#12\catcode `\^12\catcode `\_12\catcode `\%12\relax}%
\providecommand \@@startlink[1]{}%
\providecommand \@@endlink[0]{}%
\providecommand \url  [0]{\begingroup\@sanitize@url \@url }%
\providecommand \@url [1]{\endgroup\@href {#1}{\urlprefix }}%
\providecommand \urlprefix  [0]{URL }%
\providecommand \Eprint [0]{\href }%
\providecommand \doibase [0]{http://dx.doi.org/}%
\providecommand \selectlanguage [0]{\@gobble}%
\providecommand \bibinfo  [0]{\@secondoftwo}%
\providecommand \bibfield  [0]{\@secondoftwo}%
\providecommand \translation [1]{[#1]}%
\providecommand \BibitemOpen [0]{}%
\providecommand \bibitemStop [0]{}%
\providecommand \bibitemNoStop [0]{.\EOS\space}%
\providecommand \EOS [0]{\spacefactor3000\relax}%
\providecommand \BibitemShut  [1]{\csname bibitem#1\endcsname}%
\let\auto@bib@innerbib\@empty
\bibitem [{\citenamefont {Tinkham}(1996)}]{Tinkham}%
  \BibitemOpen
  \bibfield  {author} {\bibinfo {author} {\bibfnamefont {M.}~\bibnamefont
  {Tinkham}},\ }\href@noop {} {\emph {\bibinfo {title} {Introduction to
  Superconductivity}}}\ (\bibinfo  {publisher} {New York: Dover},\ \bibinfo
  {year} {1996})\BibitemShut {NoStop}%
\bibitem [{\citenamefont {Gurevich}(2006)}]{Gurevich2006}%
  \BibitemOpen
  \bibfield  {author} {\bibinfo {author} {\bibfnamefont {A.}~\bibnamefont
  {Gurevich}},\ }\href {\doibase 10.1063/1.2162264} {\bibfield  {journal}
  {\bibinfo  {journal} {App. Phys. Lett.}\ }\textbf {\bibinfo {volume} {88}},\
  \bibinfo {pages} {012511} (\bibinfo {year} {2006})}\BibitemShut {NoStop}%
\bibitem [{\citenamefont {Pavese}(1998)}]{Pavese1998}%
  \BibitemOpen
  \bibfield  {author} {\bibinfo {author} {\bibfnamefont {F.}~\bibnamefont
  {Pavese}},\ }in\ \href {\doibase 10.1201/9781420050271.chg10} {\emph
  {\bibinfo {booktitle} {Handbook of Applied Superconductivity}}},\
  Vol.~\bibinfo {volume} {2},\ \bibinfo {editor} {edited by\ \bibinfo {editor}
  {\bibfnamefont {B.}~\bibnamefont {Seeber}}}\ (\bibinfo  {publisher} {Taylor
  \& Francis},\ \bibinfo {year} {1998})\ p.\ \bibinfo {pages}
  {1461–1483}\BibitemShut {NoStop}%
\bibitem [{\citenamefont {Denis}(2007)}]{DenisPhD}%
  \BibitemOpen
  \bibfield  {author} {\bibinfo {author} {\bibfnamefont {S.}~\bibnamefont
  {Denis}},\ }\href@noop {} {\emph {\bibinfo {title} {Magnetic shielding with
  high-temperature superconductors}}},\ PhD Thesis, University of Li\`{e}ge\
  (\bibinfo {year} {2007})\BibitemShut {NoStop}%
\bibitem [{\citenamefont {Claycomb}\ and\ \citenamefont
  {J.~H.~Miller}(1999)}]{claycomb1999}%
  \BibitemOpen
  \bibfield  {author} {\bibinfo {author} {\bibfnamefont {J.~R.}\ \bibnamefont
  {Claycomb}}\ and\ \bibinfo {author} {\bibfnamefont {J.}~\bibnamefont
  {J.~H.~Miller}},\ }\href {\doibase 10.1063/1.1150113} {\bibfield  {journal}
  {\bibinfo  {journal} {Review of Scientific Instruments}\ }\textbf {\bibinfo
  {volume} {70}},\ \bibinfo {pages} {4562} (\bibinfo {year}
  {1999})}\BibitemShut {NoStop}%
\bibitem [{\citenamefont {Obi}\ \emph {et~al.}(2004)\citenamefont {Obi},
  \citenamefont {Ikebe},\ and\ \citenamefont {Fujishiro}}]{Obi2004}%
  \BibitemOpen
  \bibfield  {author} {\bibinfo {author} {\bibfnamefont {Y.}~\bibnamefont
  {Obi}}, \bibinfo {author} {\bibfnamefont {M.}~\bibnamefont {Ikebe}}, \ and\
  \bibinfo {author} {\bibfnamefont {H.}~\bibnamefont {Fujishiro}},\ }\href
  {\doibase 10.1023/B:JOLT.0000044239.59812.06} {\bibfield  {journal} {\bibinfo
   {journal} {Journal of Low Temperature Physics}\ }\textbf {\bibinfo {volume}
  {137}},\ \bibinfo {pages} {125} (\bibinfo {year} {2004})}\BibitemShut
  {NoStop}%
\bibitem [{\citenamefont {{c}ek}\ \emph {et~al.}(1996)\citenamefont {{c}ek},
  \citenamefont {Pollert},\ and\ \citenamefont
  {Hejtm\'{a}nek}}]{plechacek1996}%
  \BibitemOpen
  \bibfield  {author} {\bibinfo {author} {\bibfnamefont {V.~P.}\ \bibnamefont
  {{c}ek}}, \bibinfo {author} {\bibfnamefont {E.}~\bibnamefont {Pollert}}, \
  and\ \bibinfo {author} {\bibfnamefont {J.}~\bibnamefont {Hejtm\'{a}nek}},\
  }\href {\doibase http://dx.doi.org/10.1016/0254-0584(95)01627-7} {\bibfield
  {journal} {\bibinfo  {journal} {Materials Chemistry and Physics}\ }\textbf
  {\bibinfo {volume} {43}},\ \bibinfo {pages} {95 } (\bibinfo {year}
  {1996})}\BibitemShut {NoStop}%
\bibitem [{\citenamefont {Gurevich}(2015)}]{Gurevich2015}%
  \BibitemOpen
  \bibfield  {author} {\bibinfo {author} {\bibfnamefont {A.}~\bibnamefont
  {Gurevich}},\ }\href {\doibase 10.1063/1.4905711} {\bibfield  {journal}
  {\bibinfo  {journal} {AIP Adv.}\ }\textbf {\bibinfo {volume} {5}},\ \bibinfo
  {pages} {017112} (\bibinfo {year} {2015})}\BibitemShut {NoStop}%
\bibitem [{\citenamefont {Posen}\ \emph {et~al.}(2013)\citenamefont {Posen},
  \citenamefont {Catelani},\ and\ \citenamefont {Liepe}}]{SRF13}%
  \BibitemOpen
  \bibfield  {author} {\bibinfo {author} {\bibfnamefont {S.}~\bibnamefont
  {Posen}}, \bibinfo {author} {\bibfnamefont {G.}~\bibnamefont {Catelani}}, \
  and\ \bibinfo {author} {\bibfnamefont {M.}~\bibnamefont {Liepe}},\ }in\ \href
  {http://arxiv.org/abs/1309.3239} {\emph {\bibinfo {booktitle} {Proceedings of
  the Sixteenth Conference on RF Superconductivity}}}\ (\bibinfo {address}
  {Paris},\ \bibinfo {year} {2013})\BibitemShut {NoStop}%
\bibitem [{\citenamefont {Liepe}\ and\ \citenamefont
  {Posen}(2013)}]{liepe2013}%
  \BibitemOpen
  \bibfield  {author} {\bibinfo {author} {\bibfnamefont {M.}~\bibnamefont
  {Liepe}}\ and\ \bibinfo {author} {\bibfnamefont {S.}~\bibnamefont {Posen}}\
  }(\bibinfo  {publisher} {Presented at the SRF Conference, Paris, France},\
  \bibinfo {year} {2013})\BibitemShut {NoStop}%
\bibitem [{\citenamefont {Valles}(2014)}]{Valles2014}%
  \BibitemOpen
  \bibfield  {author} {\bibinfo {author} {\bibfnamefont {N.~R.}\ \bibnamefont
  {Valles}},\ }\emph {\bibinfo {title} {{Pushing the Frontiers of
  Superconducting Radio Frequency Science: From the Temperature Dependence of
  the Superheating Field of Niobium to Higher-Order Mode Damping in Very High
  Quality Factor Accelerating Structures}}},\ \href@noop {} {Ph.D. thesis},\
  \bibinfo  {school} {Cornell University} (\bibinfo {year} {2014})\BibitemShut
  {NoStop}%
\bibitem [{\citenamefont {Saint-James}(1965)}]{SaintJames1965}%
  \BibitemOpen
  \bibfield  {author} {\bibinfo {author} {\bibfnamefont {D.}~\bibnamefont
  {Saint-James}},\ }\href {\doibase
  http://dx.doi.org/10.1016/0031-9163(65)90810-3} {\bibfield  {journal}
  {\bibinfo  {journal} {Physics Letters}\ }\textbf {\bibinfo {volume} {16}},\
  \bibinfo {pages} {218 } (\bibinfo {year} {1965})}\BibitemShut {NoStop}%
\bibitem [{\citenamefont {Fink}(1969)}]{Fink1969}%
  \BibitemOpen
  \bibfield  {author} {\bibinfo {author} {\bibfnamefont {H.~J.}\ \bibnamefont
  {Fink}},\ }\href {\doibase 10.1103/PhysRev.177.732} {\bibfield  {journal}
  {\bibinfo  {journal} {Phys. Rev.}\ }\textbf {\bibinfo {volume} {177}},\
  \bibinfo {pages} {732} (\bibinfo {year} {1969})}\BibitemShut {NoStop}%
\bibitem [{\citenamefont {Stejic}\ \emph {et~al.}(1994)\citenamefont {Stejic},
  \citenamefont {Gurevich}, \citenamefont {Kadyrov},\ and\ \citenamefont
  {Christen}}]{Stejic1994}%
  \BibitemOpen
  \bibfield  {author} {\bibinfo {author} {\bibfnamefont {G.}~\bibnamefont
  {Stejic}}, \bibinfo {author} {\bibfnamefont {A.}~\bibnamefont {Gurevich}},
  \bibinfo {author} {\bibfnamefont {E.}~\bibnamefont {Kadyrov}}, \ and\
  \bibinfo {author} {\bibfnamefont {D.}~\bibnamefont {Christen}},\ }\href
  {http://prb.aps.org/abstract/PRB/v49/i2/p1274\_1} {\bibfield  {journal}
  {\bibinfo  {journal} {Physical Review B}\ }\textbf {\bibinfo {volume} {49}}
  (\bibinfo {year} {1994})}\BibitemShut {NoStop}%
\bibitem [{\citenamefont {Tinkham}(2004)}]{Tinkham2004}%
  \BibitemOpen
  \bibfield  {author} {\bibinfo {author} {\bibfnamefont {M.}~\bibnamefont
  {Tinkham}},\ }\href
  {http://books.google.com/books/about/Introduction\_to\_Superconductivity.html?id=k6AO9nRYbioC\&pgis=1}
  {\emph {\bibinfo {title} {{Introduction to Superconductivity}}}}\ (\bibinfo
  {publisher} {Dover Publications},\ \bibinfo {address} {New York},\ \bibinfo
  {year} {2004})\ p.\ \bibinfo {pages} {454}\BibitemShut {NoStop}%
\bibitem [{\citenamefont {Orlando}\ \emph {et~al.}(1979)\citenamefont
  {Orlando}, \citenamefont {McNiff}, \citenamefont {Foner},\ and\ \citenamefont
  {Beasley}}]{Orlando1979}%
  \BibitemOpen
  \bibfield  {author} {\bibinfo {author} {\bibfnamefont {T.~P.}\ \bibnamefont
  {Orlando}}, \bibinfo {author} {\bibfnamefont {E.~J.}\ \bibnamefont {McNiff}},
  \bibinfo {author} {\bibfnamefont {S.}~\bibnamefont {Foner}}, \ and\ \bibinfo
  {author} {\bibfnamefont {M.~R.}\ \bibnamefont {Beasley}},\ }\href {\doibase
  10.1103/PhysRevB.19.4545} {\bibfield  {journal} {\bibinfo  {journal} {Phys.
  Rev. B}\ }\textbf {\bibinfo {volume} {19}},\ \bibinfo {pages} {4545}
  (\bibinfo {year} {1979})}\BibitemShut {NoStop}%
\bibitem [{\citenamefont {Hein}(1999)}]{Hein1999}%
  \BibitemOpen
  \bibfield  {author} {\bibinfo {author} {\bibfnamefont {M.}~\bibnamefont
  {Hein}},\ }\href {http://books.google.com/books?id=5MIDpYI1z20C\&pgis=1}
  {\emph {\bibinfo {title} {{High-Temperature-Superconductor Thin Films at
  Microwave Frequencies}}}}\ (\bibinfo  {publisher} {Springer},\ \bibinfo
  {address} {New York},\ \bibinfo {year} {1999})\BibitemShut {NoStop}%
\bibitem [{\citenamefont {Harden}\ and\ \citenamefont
  {Arp}(1963)}]{Harden1963}%
  \BibitemOpen
  \bibfield  {author} {\bibinfo {author} {\bibfnamefont {J.}~\bibnamefont
  {Harden}}\ and\ \bibinfo {author} {\bibfnamefont {V.}~\bibnamefont {Arp}},\
  }\href {\doibase 10.1016/0011-2275(63)90029-8} {\bibfield  {journal}
  {\bibinfo  {journal} {Cryogenics (Guildf).}\ }\textbf {\bibinfo {volume}
  {3}},\ \bibinfo {pages} {105} (\bibinfo {year} {1963})}\BibitemShut {NoStop}%
\bibitem [{\citenamefont {Transtrum}\ \emph {et~al.}(2011)\citenamefont
  {Transtrum}, \citenamefont {Catelani},\ and\ \citenamefont
  {Sethna}}]{Transtrum2011}%
  \BibitemOpen
  \bibfield  {author} {\bibinfo {author} {\bibfnamefont {M.~K.}\ \bibnamefont
  {Transtrum}}, \bibinfo {author} {\bibfnamefont {G.}~\bibnamefont {Catelani}},
  \ and\ \bibinfo {author} {\bibfnamefont {J.~P.}\ \bibnamefont {Sethna}},\
  }\href {\doibase 10.1103/PhysRevB.83.094505} {\bibfield  {journal} {\bibinfo
  {journal} {Physical Review B}\ }\textbf {\bibinfo {volume} {83}},\ \bibinfo
  {pages} {094505} (\bibinfo {year} {2011})}\BibitemShut {NoStop}%
\bibitem [{\citenamefont {Maxfield}\ and\ \citenamefont
  {McLean}(1965)}]{Maxfield1965}%
  \BibitemOpen
  \bibfield  {author} {\bibinfo {author} {\bibfnamefont {B.~W.}\ \bibnamefont
  {Maxfield}}\ and\ \bibinfo {author} {\bibfnamefont {W.~L.}\ \bibnamefont
  {McLean}},\ }\href {\doibase 10.1103/PhysRev.139.A1515} {\bibfield  {journal}
  {\bibinfo  {journal} {Phys. Rev.}\ }\textbf {\bibinfo {volume} {139}},\
  \bibinfo {pages} {A1515} (\bibinfo {year} {1965})}\BibitemShut {NoStop}%
\bibitem [{\citenamefont {Hein}\ \emph {et~al.}(2001)\citenamefont {Hein},
  \citenamefont {Perpeet},\ and\ \citenamefont {Muller}}]{Hein2001}%
  \BibitemOpen
  \bibfield  {author} {\bibinfo {author} {\bibfnamefont {M.}~\bibnamefont
  {Hein}}, \bibinfo {author} {\bibfnamefont {M.}~\bibnamefont {Perpeet}}, \
  and\ \bibinfo {author} {\bibfnamefont {G.}~\bibnamefont {Muller}},\ }\href
  {\doibase 10.1109/77.919801} {\bibfield  {journal} {\bibinfo  {journal} {IEEE
  Transactions on Appiled Superconductivity}\ }\textbf {\bibinfo {volume}
  {11}},\ \bibinfo {pages} {3434} (\bibinfo {year} {2001})}\BibitemShut
  {NoStop}%
\bibitem [{\citenamefont {Oates}\ \emph {et~al.}(1991)\citenamefont {Oates},
  \citenamefont {Anderson}, \citenamefont {Chin}, \citenamefont {Derov},
  \citenamefont {Dresselhaus},\ and\ \citenamefont {Dresselhaus}}]{Oates1991}%
  \BibitemOpen
  \bibfield  {author} {\bibinfo {author} {\bibfnamefont {D.~E.}\ \bibnamefont
  {Oates}}, \bibinfo {author} {\bibfnamefont {A.~C.}\ \bibnamefont {Anderson}},
  \bibinfo {author} {\bibfnamefont {C.~C.}\ \bibnamefont {Chin}}, \bibinfo
  {author} {\bibfnamefont {J.~S.}\ \bibnamefont {Derov}}, \bibinfo {author}
  {\bibfnamefont {G.}~\bibnamefont {Dresselhaus}}, \ and\ \bibinfo {author}
  {\bibfnamefont {M.~S.}\ \bibnamefont {Dresselhaus}},\ }\href {\doibase
  10.1103/PhysRevB.43.7655} {\bibfield  {journal} {\bibinfo  {journal} {Phys.
  Rev. B}\ }\textbf {\bibinfo {volume} {43}},\ \bibinfo {pages} {7655}
  (\bibinfo {year} {1991})}\BibitemShut {NoStop}%
\bibitem [{\citenamefont {Wang}\ \emph {et~al.}(2001)\citenamefont {Wang},
  \citenamefont {Plackowski},\ and\ \citenamefont {Junod}}]{Wang2001}%
  \BibitemOpen
  \bibfield  {author} {\bibinfo {author} {\bibfnamefont {Y.}~\bibnamefont
  {Wang}}, \bibinfo {author} {\bibfnamefont {T.}~\bibnamefont {Plackowski}}, \
  and\ \bibinfo {author} {\bibfnamefont {A.}~\bibnamefont {Junod}},\ }\href
  {\doibase http://dx.doi.org/10.1016/S0921-4534(01)00617-7} {\bibfield
  {journal} {\bibinfo  {journal} {Physica C: Superconductivity}\ }\textbf
  {\bibinfo {volume} {355}},\ \bibinfo {pages} {179 } (\bibinfo {year}
  {2001})}\BibitemShut {NoStop}%
\bibitem [{\citenamefont {Shmidt}(1972)}]{Shmidt1972}%
  \BibitemOpen
  \bibfield  {author} {\bibinfo {author} {\bibfnamefont {V.}~\bibnamefont
  {Shmidt}},\ }\href {http://www.jetp.ac.ru/cgi-bin/dn/e\_034\_01\_0211.pdf}
  {\bibfield  {journal} {\bibinfo  {journal} {Soviet Physics JETP}\ }\textbf
  {\bibinfo {volume} {34}} (\bibinfo {year} {1972})}\BibitemShut {NoStop}%
\bibitem [{\citenamefont {Bean}\ and\ \citenamefont
  {Livingston}(1964)}]{Bean1964}%
  \BibitemOpen
  \bibfield  {author} {\bibinfo {author} {\bibfnamefont {C.}~\bibnamefont
  {Bean}}\ and\ \bibinfo {author} {\bibfnamefont {J.}~\bibnamefont
  {Livingston}},\ }\href
  {http://www.osti.gov/energycitations/product.biblio.jsp?osti\_id=4080469}
  {\bibfield  {journal} {\bibinfo  {journal} {Phys. Rev. Letters}\ }\textbf
  {\bibinfo {volume} {12}},\ \bibinfo {pages} {4} (\bibinfo {year}
  {1964})}\BibitemShut {NoStop}%
\bibitem [{\citenamefont {Catelani}\ and\ \citenamefont
  {Sethna}(2008)}]{Catelani2008}%
  \BibitemOpen
  \bibfield  {author} {\bibinfo {author} {\bibfnamefont {G.}~\bibnamefont
  {Catelani}}\ and\ \bibinfo {author} {\bibfnamefont {J.}~\bibnamefont
  {Sethna}},\ }\href {\doibase 10.1103/PhysRevB.78.224509} {\bibfield
  {journal} {\bibinfo  {journal} {Physical Review B}\ }\textbf {\bibinfo
  {volume} {78}},\ \bibinfo {pages} {224509} (\bibinfo {year}
  {2008})}\BibitemShut {NoStop}%
\bibitem [{\citenamefont {Lin}\ and\ \citenamefont {Gurevich}(2012)}]{Lin2012}%
  \BibitemOpen
  \bibfield  {author} {\bibinfo {author} {\bibfnamefont {F.~P.-J.}\
  \bibnamefont {Lin}}\ and\ \bibinfo {author} {\bibfnamefont {a.}~\bibnamefont
  {Gurevich}},\ }\href {\doibase 10.1103/PhysRevB.85.054513} {\bibfield
  {journal} {\bibinfo  {journal} {Phys. Rev. B}\ }\textbf {\bibinfo {volume}
  {85}},\ \bibinfo {pages} {054513} (\bibinfo {year} {2012})}\BibitemShut
  {NoStop}%
\end{thebibliography}%
\end{document}